\documentclass[a4paper,11pt]{article}
\pdfoutput=1 
\usepackage{jheppub} 
\usepackage{bbm,bm,graphicx,mathtools,color,hyperref,slashed}
\usepackage{amssymb,amsmath}
\usepackage[T1]{fontenc}

\usepackage[all]{xy}

\graphicspath{{./Figures/}}

\newcommand{\be}{\begin{equation}}      
\newcommand{\ee}{\end{equation}}      
\newcommand{\bea}{\begin{eqnarray}}      
\newcommand{\eea}{\end{eqnarray}}
\newcommand{\bi}{\begin{itemize}}
\newcommand{\ei}{\end{itemize}}

\newcommand{\im}{\mathrm{i}}
\newcommand{\rmc}{\mathrm{c}}
\newcommand{\rme}{\mathrm{e}}
\newcommand{\rmf}{\mathrm{f}}

\newcommand{\wh}[1]{{\widehat{#1}}}

\newcommand{\diff}{\mathrm{d}}

\newcommand\diag{\operatorname{diag}}

\newcommand\Dcal{\mathcal{D}}

\newcommand\D{\mathcal{D}}
\newcommand\e{\mathrm{e}}
\newcommand\p{\partial}

\newcommand\rd{{\rm d}}

\newcommand\lm{{\lambda} }

\newcommand\ri{{\mathrm i}} 

\title{
Effective gauge theories of superfluidity with topological order
}

\author{Yuji Hirono$^{1,2}$,}
\emailAdd{yuji.hirono@apctp.org}

\affiliation[1]{
Asia Pacific Center for Theoretical Physics, Pohang, Gyeongbuk 37673, Korea
}
\affiliation[2]{
Department of Physics, POSTECH, Pohang, Gyeongbuk 37673, Korea
}

\author{Yuya Tanizaki$^3$}
\emailAdd{ytaniza@ncsu.edu}

\affiliation[3]{
Department of Physics, North Carolina State University, Raleigh, NC 27607, USA
}

\abstract{
We discuss the low-energy dynamics of superfluidity with 
topological order in $(3+1)$ spacetime dimensions. 
We generalize a topological $BF$ theory by introducing a non-square $K$ matrix, and this generalized $BF$ theory can describe massless Nambu-Goldstone bosons and anyonic statistics between vortices and quasiparticles. 
We discuss the general structure of discrete and continuous higher-form symmetries in this theory, which can be used to classify quantum phases. 
We describe how to identify the appearance of topological order in 
such systems and discuss its relation to a mixed 't~Hooft anomaly between 
discrete higher-form symmetries. 
We apply this framework to the color-flavor locked phase of dense QCD, which shows anyonic particle-vortex statistics while no topological order appears. 
An explicit example of superfluidity with topological order is discussed. 
}

\begin{document}
\maketitle
\section{Introduction}\label{sec:intro}

Classification of phases of matter has been one of the most fundamental 
problems in the physics of many-body systems.
Different phases of matter have been classified by their symmetries, 
which led to the theory of spontaneously
symmetry breaking~\cite{landau1937theory, ginzburg1950theory,Nambu:1960tm, Nambu:1961fr}.
It is now realized that quantum phases of matter
depends also on ``topology'', and Ginzburg-Landau (GL) type 
classification is not sufficient. 
An important class of such states is called topological order, and 
there are nontrivial long-range correlation even though no
massless excitations exist: topological degeneracy of ground states,
anyon statistics of quasiparticles, and so on~\cite{Wen:1989iv, PhysRevB.44.274, Wen200710, HANSSON2004497, sachdev2011quantum, Chen:2010gda, PhysRevLett.96.110405, PhysRevLett.96.110404}.
Low-energy effective description of topological order is given by
topological field theories, and the presence of ``deconfined'' dynamical
gauge fields plays an important role for those nontrivial
long-range phenomena with mass gap~\cite{Dijkgraaf:1990nc, Wen:1992uk}.
Certain classes of topological orders
can be understood as a consequences of spontaneous breaking of 
higher-form (or generalized) global symmetries~\cite{Gaiotto:2014kfa, Wen:2018zux}. 

One possible direction of further developments on quantum many-body 
physics is to understand the role of topology in the presence of
gapless degrees of freedom. 
Although we do not have a complete consensus about the definition of
topological order in gapless systems, let us temporarily consider it in
this paper as a quantum system which has deconfined gauge 
fields in addition to local gapless excitations.
Such theoretical models have been recently studied in the context of
quantum criticality of high-$T_{\rm c}$ cuprates, using gauged GL
model~\cite{Chatterjee:2017pqv, Sachdev:2017hzd, Scheurer:2017jcp,
Sachdev:2018nbk}.
There is also an example of such description in the physics of  QCD~\cite{Stephanov:2004wx, Fukushima:2010bq, Ren:2004nn, Casalbuoni:2018haw}. 
At large baryon densities, QCD matter is expected to exhibit color 
color superconductivity~\cite{Barrois:1977xd, Bailin:1983bm, Alford:1998mk, Schafer:1998ef, Alford:2007xm}, 
which has topological vortices~\cite{Balachandran:2005ev, Nakano:2007dr, Eto:2013hoa, Yamamoto:2018vgg}. 
Understanding the role of topology would be important 
in the discussion of the phase structure of dense nuclear matter \cite{Schafer:1998ef, Hirono:2018fjr, Cherman:2018jir} or the possible continuity of vortices
between a nuclear superfluid and a color superconductor \cite{Alford:2018mqj, Chatterjee:2018nxe}. 
The low-energy theory of a color superconducting phase 
is described by a topological field theory coupled with massless
Nambu-Goldstone (NG) bosons~\cite{Hirono:2018fjr, Hirono:2010gq, Cherman:2018jir}.

Motivated by these recent developments on the possibility of topological order in
gapless systems, 
we study a general framework for studying superfluidity coupled to
$BF$-type topological field theory~\cite{Bergeron:1994ym, HANSSON2004497, Cho:2010rk, Putrov:2016qdo, PhysRevB.94.045113}. 
Starting from a gauged GL model, we derive a dual gauge theory. 
The effective theory is a generalized $BF$ theory
with a non-square $K$ matrix coupled with massless NG bosons. 
The system is shown to acquire discrete and continuous 2-form/1-form symmetries. 
As a consequence of the emergent symmetries, 
the system is shown to exhibit fractional braiding statistics between
vortices and quasiparticles. 
We examine the condition when a topological order appears, 
which can be also seen as the 
existence of a mixed 't~Hooft anomaly \cite{tHooft:1979rat}
between higher-form symmetries. 

This paper is organized as follows. 
In Sec.~\ref{sec:eft}, 
we introduce a low-energy effective theory of superfluidity that 
can also have topological order. 
In Sec.~\ref{sec:emergent}, 
we identify the continuous and discrete higher-form symmetries of the system 
and discuss the braiding of quasiparticles and vortices. 
We also discuss how to detect the topological order in this theory and 
its interpretation as a mixed 't~Hooft anomaly. 
In Sec.~\ref{sec:cfl}, we describe the topological properties of the color-flavor locked phase of dense QCD as an application of the framework. 
This section is the follow-up of the previous paper~\cite{Hirono:2018fjr} with more detailed explanations. 
In Sec.~\ref{sec:tos}, we discuss an explicit example of superfluidity with topological order. 
Section~\ref{sec:summary} is devoted to a summary and outlook. 
In Appendix~\ref{app:mp}, we summarize the properties of the Moore-Penrose inverse. 
In Appendix~\ref{app:delta}, we provide a summary of the properties of delta-function forms. 
In Appendix~\ref{app:derivation}, we give a derivation of the braiding phase using the effective theory. 
Appendix~\ref{sec:smith} is devoted to a discussion about the consequence of basis changes.

\section{Effective field theory of topologically ordered superfluidity}\label{sec:eft}

We aim at describing the low-energy behavior of superfluids with topological order. 
Here, let us construct a generic low-energy effective theory that consists of $2\pi$-periodic compact scalar fields and $U(1)$ gauge fields.

\subsection{Effective Lagrangian of general Abelian-Higgs models}

We consider a $3+1$-dimensional theory with multiple $U(1)$ symmetries and 
some parts of them are gauged and couple to dynamical $U(1)$ 1-form gauge fields. 
We are interested in the low-energy regime of the theory
and in this limit, the remaining degrees of freedom are
massless modes, that are the  NG bosons associated with 
the spontaneous breaking of $U(1)$ symmetries. 
Thus, the system is a superfluid. 
In addition, there can be topological degrees of freedom. 

Let us give a derivation of an effective theory for describing such a system. 
We take variables $\phi_i$,
that are (would-be) NG modes associated with 
the breaking of $U(1)$ symmetries. These are $2\pi$-periodic scalar fields. 
They couple to gauge fields through a covariant derivative, 
\begin{equation}
 \rd_a \phi_i \equiv  \rd \phi_i + K_{iA} a_A ,
\end{equation}
and $a_A$ are dynamical Abelian 1-form gauge fields, 
$a_A(x) = (a_A)_\mu \diff x^\mu$ ($A = 1, \cdots, |A|$),
and $K_{iA}$ is a $|i| \times |A|$\footnote{
We use the notation to represent the number of rows (or columns)
of a matrix by the absolute value of the index. 
} integer-valued matrix. 
Let us call $a_A(x)$ as photons. 
The covariant derivative is invariant under the $0$-form gauge transformation,
\be
\phi_i\mapsto \phi_i-K_{iA}\lambda_A,  \quad
a_{A}\mapsto a_A+\diff \lambda_A,
\ee
where gauge parameters $\lambda_A$ are also $2\pi$-periodic scalars. 
Because of this interaction, a part of the would-be NG modes are Higgsed. 
We start with an action, 
\begin{equation}
 S =
  \frac{1}{2}
  H_{ij}\int \rd_a \phi_i \wedge \star\, \rd_a \phi_j 
 + \frac{1}{2}G^a_{AB}\int \rd a_A \wedge \star \, \rd a_B . 
\label{eq:EFT_general_01}
\end{equation}
The positivity of the kinetic terms require that $H$ and $G^a$ are positive-definite real symmetric matrices. 

We shall take an Abelian dual of this theory~\cite{Banks:2010zn}. 
The action can be rewritten by introducing $\mathbb{R}$-valued $3$-form fields $h_i$ as 
\begin{equation}
 S = \frac{1}{8\pi^2}H^{-1}_{ij} \int h_i \wedge \star\, h_j
  - \frac{\im}{2\pi} \int h_i \wedge \rd_a \phi_i  
	+ \frac{1}{2}G^a_{AB}\int \rd a_A \wedge \star \, \rd a_B ,
\label{eq:EFT_general_02}
\end{equation}
Solving the equation of motion (EOM) for $h_i$, we obtain $h_i = 2\pi \im H_{ij}\star\, \rd_a \phi_j$ and get the original action (\ref{eq:EFT_general_01}). 
Solving the EOM for $\phi_i$, instead, we find $\rd h_i = 0$, and it can be solved as 
\begin{equation}
 h_i = \rd b_i  ,
\label{eq:dual_gauge_field}
\end{equation}
where
$b_i(x) = \frac{1}2 (b_i)_{\mu\nu} dx^\mu \wedge
dx^\nu$ ($i = 1, \cdots, |i|$) are $2$-form $U(1)$ gauge fields\footnote{This normalization is determined by the global structure: Let us set our spacetime as $4$-torus $T^4$ of size $L$ as an example, then the $2\pi$-periodic scalar $\phi$ can be decomposed as $\phi={2\pi\over L}n_{\mu}x^{\mu}+\tilde{\phi}$, where $n_{\mu}\in\mathbb{Z}$ and $\tilde{\phi}$ is the $\mathbb{R}$-valued field. The above EOM, $\rd h=0$, comes out of the path integral over $\tilde{\phi}$. The summation over $\{n_{\mu}\}\in\mathbb{Z}^4$ further requires that $\int_{T^3}h \in 2\pi \mathbb{Z}$ for each $3$-torus and we find the correct normalization (\ref{eq:dual_gauge_field}). }. 
Plugging this into $h_i$, the action is rewritten as 
\begin{equation}
 S = \frac{1}{2} G^b_{ij}
 \int \rd b_i \wedge \star \, \rd b_j 
  + \frac{1}{2}G^a_{AB} \int \rd a_A \wedge \star \rd a_B
  + \im \frac{K_{iA}}{2\pi} \int b_i \wedge \rd a_A, 
\label{eq:EFT_general_03}
\end{equation}
where we introduce
$
G^b_{ij} \equiv H^{-1}_{ij} / 4 \pi^2 . 
$
The first two terms are the usual kinetic terms, and the last term is the topological $BF$ term\footnote{One could consider further generalization by adding $a_A\wedge a_B\wedge \diff a_C$, $b_i\wedge b_j$, etc., with appropriate coefficients and modification of gauge transformations, like a twist term of Dijkgraaf-Witten theory~\cite{Dijkgraaf:1990nc, Kapustin:2014gua, Tiwari:2016zru}. In this paper, however, we do not pursue along this direction.}. 
Physical observables can be calculated
by the partition function, 
\begin{equation}
 Z = \int \Dcal a \Dcal b \, \e^{ - S[a,b]} . 
\end{equation}
In this path integral, we sum over all possible gauge fields, satisfying the canonical Dirac quantization conditions, 
\begin{equation}
\int_{S} \diff a_A\in 2\pi\mathbb{Z},\; \int_{V} \diff b_i\in 2\pi\mathbb{Z}, 
\label{eq:normalization_gauge_01}
\end{equation}
for each closed $2$-submanifold $S$ and $3$-submanifold $V$ of the spacetime.

Physical operators that we will focus on are 
Wilson loop operators and vortex operators, 
\begin{equation}
 W_{A}(C) = \exp \left( \ri \int_C a_A \right),
  \quad
 V_{i}(S) = \exp \left( \ri  \int_S b_i \right),
\label{eq:Wilson_loop_surface} 
\end{equation}
where $C$ is a world-line of a test particle, and $S$ is a vortex world-sheet.

\subsection{Classification of spectra}\label{sec:local_dynamics}

Since we are interested in the low-energy physics,
we only retain massless modes and topological sector. 
Before studying the global nature of the theory, let us clarify its local dynamics
to identify the massless sector of $b$ and $a$. 
For this purpose, let us write down the EOMs of $b_i$ and $a_A$:
\begin{eqnarray}
&&G^b_{ij} \diff \star \diff b_j-{\im\over 2\pi}K_{iA}\diff a_A=0,\nonumber\\
&&G^{a}_{AB}\diff \star \diff a_B+{\im\over 2\pi}K_{iA} \diff b_i=0. 
\end{eqnarray}
Combining these EOMs, we find that $\diff b_i$ must satisfy
\begin{equation}
\left(G^b \Delta+{1\over 4\pi^2}K (G^a)^{-1}K^T\right) \diff b=0,
\end{equation}
and that $\diff a_A$ must satisfy
\begin{equation}
\left(G^a \Delta+{1\over 4\pi^2}K^T (G^b)^{-1} K\right)\diff a=0. 
\end{equation}
Here, $\Delta=\diff \delta+\delta \diff$ is the form Laplacian with the codifferential $\delta=-\star \diff \star$, and $K^T$ represents the transpose of $K$ matrix. 
Therefore, the mass matrix $M^b$ of $b_i$ is given by 
\begin{equation}
(M^b)^2={1\over 4\pi^2}(G^b)^{-1/2} K (G^a)^{-1} K^T (G^b)^{-1/2}, 
\end{equation}
and the mass matrix $M^a$ of $a_A$ is given by 
\begin{equation}
(M^a)^2={1\over 4\pi^2}(G^a)^{-1/2} K^T (G^b)^{-1} K (G^a)^{-1/2}. 
\end{equation}
Here, we note that squared roots of $G^a$ and $G^b$ are well defined since they are positive matrix.

Let us discuss how massless degrees of freedom depend on the structure of the $K$ matrix. 
When $|i| = |A|$ (i.e. $K$ is a square matrix) and $\det K \neq 0$, all the particles get nonzero mass because neither $M^b$ nor $M^a$ have zero eigenvalues, 
and the $BF$ theory for superconductivity is reproduced. 
We are interested in the situation where superfluidity is present. 
In this case, there exists at least one massless NG modes, 
which is realized when ${\rm dim} ({\rm coker}\,K) \neq 0$.
Indeed, for each vector $\bm{D}^{\bar\alpha}\in \mathrm{coker}\,\, K$, i.e. 
\begin{equation}
(\bm{D}^{\bar\alpha})^T\cdot K=0,
\end{equation}
we can find the null eigenvector of the mass matrix:
\begin{equation}
(M^b)^2 \sqrt{G^b} \bm{D}^{\bar\alpha}=0. 
\end{equation}
Since $K$ is in general a non-square matrix, there always exist massless NG modes if $|i|>|A|$.  

There can also be remaining massless photons, 
when ${\rm dim} ({\rm ker}\, K) \neq 0$. 
We denote the basis of the kernel and cokernel as 
$\bm C^{\alpha} \in {\rm ker} \,\, K$, and 
$\bm D^{\bar\alpha} \in {\rm coker}\,\, K$, namely, they satisfy 
\begin{equation}
 K_{iA} C^{\alpha}_A = 0,
  \quad 
 D^{\bar\alpha}_i K_{iA} = 0. 
\end{equation}
We can identify the massless NG modes and massless photons as 
\begin{equation}
 b_0 \in {\rm coker}\,\, K,
  \quad
 a_0 \in {\rm ker}\,\, K. 
\end{equation}
because $a_0$ and $b_0$ does appear in the $BF$ term. 
The numbers of massless NG modes and massless photons,
$|\bar\alpha|$ and $|\alpha|$, 
are given by the dimensions of the cokernel and kernel of $K$, respectively:
\begin{equation}
 |\bar\alpha| = {\rm dim\,}({\rm coker}\, K), 
  \quad
 |\alpha| = {\rm dim\,} ({\rm ker}\, K). 
\end{equation}
Those massless modes can be identified by projection matrices, 
\begin{equation}
(a_0)_A = P^a_{AB}\, a_B , 
\quad
(b_0)_i = P^b_{ij}\, b_j, 
\end{equation}
where $P^a$ and $P^b$ are orthogonal projectors to the kernel and
cokernel of $K$.
They can be expressed using the Moore-Penrose inverse\footnote{
We summarize the properties of the
Moore-Penrose inverse in Appendix \ref{app:mp}. 
} $K^+$
of the $K$ matrix, which is a generalization of a matrix inverse. 
Given an arbitrary matrix, the Moore-Penrose inverse always exists and is unique. 
The projectors are given by \footnote{
We use the notation where contracted matrix indices may be omitted 
when there is no confusion, for example, 
$[K^+ K]_{AB} = K^+_{Ai} K_{iB}$. 
}
\begin{equation}
 P^a_{AB} =  \delta_{AB} -  [K^+ K]_{AB} ,
  \quad
 P^b_{ij} =  \delta_{ij} - [K K^+]_{ij}. 
\end{equation}
Using this orthonormal projection, we denote the gauge fields as 
\begin{equation}
a=a_0+a_\perp, 
\quad 
b=b_0+b_\perp. 
\end{equation}
This decomposition diagonalizes the mass term, but it does not necessarily diagonalize the kinetic term: For example, the kinetic term of $b$ becomes 
\begin{equation}
{1\over 2}\int \diff b^T\wedge G^b\star\diff b={1\over 2}\int \left(\diff b_0^T\wedge G^b\star\diff b_0+2\diff b_\perp^T\wedge G^b\star\diff b_0+\diff b_\perp^T\wedge G^b\star\diff b_\perp\right). 
\end{equation}
In the low-energy limit, the last term can be neglected since it only describes the exponential decay of massive excitations, and one should retain the first and the second terms. The mixed kinetic term $\diff b_\perp^T\wedge G^b\star\diff b_0$ vanishes identically if and only if
\begin{equation}
P^b G^b (1-P^b)=0. 
\end{equation}
Since $G^b$ and $P^b$ are both symmetric matrices, this condition is equivalent to 
\begin{equation}
[G^b,P^b]=0. 
\end{equation}
We obtain the same conclusion also for the photon fields $a$. 
In the rest of this paper, we assume that
\begin{equation}
[G^a, P^a]=0,\quad  
[G^b, P^b]=0, \label{eq:assumption_orthogonality}
\end{equation}
so that the mixed kinetic terms between massless and heavy modes vanish identically\footnote{
In Appendix~\ref{sec:smith}, we discuss the consequence of the mixed kinetic terms. 
}. 
In concrete examples, the condition (\ref{eq:assumption_orthogonality}) may be implied by a certain symmetry of the UV theory. 
We expect that this condition is important to protect the topological order under the existence of gapless excitations. 
As a consequence of assumption (\ref{eq:assumption_orthogonality}), we obtain the low-energy effective action as 
\begin{equation}
 S_{\rm eff} =
  \im \frac{ K_{i A}}{2\pi} \int
 b_{i} \wedge \rd a_{A}
  + \frac{1}{2}G^b_{ij} \int  \rd (b_0)_i \wedge \star \,\rd (b_0)_j
  + \frac{1}{2}G^a_{AB} \int  \rd (a_0)_A \wedge \star \,\rd (a_0)_B,
  \label{eq:seff}
\end{equation}
where $b_0$ and $a_0$ are massless contributions as identified above. 
We call this as the generalized $BF$ theory. 

This action (\ref{eq:seff}) can describe various physical situations,
depending on the choice of $K$ matrices.
Possible physical situations are classified according to
the numbers, $|\bar\alpha|$, $|\alpha|$, $|i|$, and $|A|$.
Note that, according to Fredholm's index theorem, those numbers are
related as 
\begin{equation}
 |\bar \alpha|  - |\alpha| = |i| -|A| . 
\end{equation}
Based on this relation, possible situations of (\ref{eq:seff}) can be classified as follows: 
\begin{itemize}
 \item $|\bar\alpha| =|\alpha| = 0$, $|i| = |A|$: All the excitations
       are massive. In this case, The $K$ matrix is square and regular,
       which corresponds to $BF$ theoretical description of superconductors. 
 \item $|\alpha| = |A|=0$, $|\bar\alpha| = |i|$: Superfluids with no
       topological order. 
 \item $|\bar\alpha| = |i| = 0$, $|\alpha| = |A|$: Pure Maxwell theory. 
 \item In other cases, superfluidity and topological order may coexist. 
\end{itemize}

\subsection{Non-canonical normalization of gauge fields}

\subsubsection{Generalization to non-canonical normalization}

So far, we are working on the theory (\ref{eq:EFT_general_03}) with the canonically normalized gauge fields $b_i$ and $a_A$ as in (\ref{eq:normalization_gauge_01}). 
Instead, we can work on more general normalization of these gauge fields, and let us discuss such cases in this section. 
The motivation for this generalization is that gauge fields of the low-energy effective theory can be emergent and is not necessarily ensured to be canonically normalized when we derive it from the UV theory. 
Therefore, it is important to establish the way to analyze such cases. 
The normalization of gauge fields does not affect local dynamics, and thus the discussion in Sec.~\ref{sec:local_dynamics} is unaffected while the global nature of the theory can be changed drastically. 

The normalization of gauge fields is related to the choice of physically observable Wilson loops and vortex operators. 
Let us replace (\ref{eq:normalization_gauge_01}) by a generic normalization condition 
(Dirac quantization condition)
for $a_A$ and $b_i$ as 
\be
Q_{A B}\int_S \rd a_{B} 
\in 2\pi \mathbb Z , 
\quad 
\int_V \rd b_{j} R_{ji} 
\in 2\pi \mathbb Z , 
\label{eq:norm}
\ee
where 
$S$ and $V$ are 2D and 3D subspace without boundary, 
and $Q_{AB}$ and $R_{ij}$ are invertible matrices with integer elements. 
Correspondingly, 
the matrices $Q$ and $R$ specify the set of generators of 
gauge-invariant Wilson loops and vortex operators, 
\be
W^{(Q)}_A \equiv  
\exp 
\left(
\im Q_{AB} \int_C a_B 
\right), 
\quad 
V^{(R)}_i \equiv 
\exp 
\left(
\im  \int_S b_j \, R_{ji}
\right). 
\label{eq:wvqr}
\ee
The naive Wilson loop and surface operators (\ref{eq:Wilson_loop_surface}) are no longer gauge invariant in this normalization. 

Let us discuss the gauge redundancy  of the action. 
It is invariant under $0$-form and $1$-form gauge transformations, 
\begin{equation}
 a_A \mapsto a_A + Q^{-1}_{AB} \, \rd \lambda_B ,
  \quad 
 b_i \mapsto b_i + R^{-1}_{ji}\, \rd \lambda_j . 
\end{equation}
where $\lm_A$ is a $2\pi$-periodic scalar and $\lm_i$ is a $U(1)$ $1$-form fields. 
They satisfy 
\be
\int \rd \lambda_A \in 2\pi \mathbb Z, 
\quad 
\int \rd \lambda_i \in 2\pi \mathbb Z, 
\ee
where the integrations are over closed submanifolds of corresponding dimensions. 
Introducing the matrices $Q^{-1}$ and $R^{-1}$ is necessary 
for the consistency with the normalization condition (\ref{eq:norm}). 
The variation of the action is 
\begin{equation}
  \delta_{1} S_{\rm eff} = 
  \frac{ K_{iA} Q^{-1}_{AB}}{2\pi}
  \int \rd b_{i} \wedge \rd \lm_{B} ,
\quad 
 \delta_{2} S_{\rm eff} = \frac{R^{-1}_{ji} K_{iA}}{2\pi}
  \int  \rd \lambda_{i} \wedge \rd a_{A} . 
\end{equation}
Note that the integration gives 
\be
\int \rd b_{i} \wedge \rd \lm_{B}  = (2\pi)^2 R^{-1}_{ji} n_j m_{B}
, \quad 
  \int  \rd \lambda_{i} \wedge \rd a_{A} 
  = (2\pi)^2 Q^{-1}_{AB} m'_B n'_i 
\ee
where $n_i, n_i',m_B, m'_B$ are integer vectors.
The change of the action under 1-form and 2-form gauge transformation 
is now given by 
\begin{equation}
  \delta_{1} S_{\rm eff} = 
  2\pi  
  \, 
  n^T R^{-1} K Q^{-1}m
, 
\, \quad 
 \delta_{2} S_{\rm eff} = 
  2\pi \,
   n'^{T} R^{-1} K Q^{-1}m'
   \label{eq:gauge-delta}
\end{equation}
Thus, the gauge invariance requires that the 
$K$ matrix should be chosen so that the each element of 
the matrix $[R^{-1} K Q^{-1}]_{iA}$ is an integer. This reproduces the fact that $K$ should be an integer-valued matrix when we take $Q=1$ and $R=1$, i.e., the canonical normalization~(\ref{eq:normalization_gauge_01}) of gauge fields.

\subsubsection{Basis change of gauge fields}\label{sec:basis_change}

As we have shown, a theory can be specified by a set of integer-valued matrices $(K,Q,R)$. 
However, not every theory associated with a set $(K,Q,R)$ describes a distinct system. 
This ambiguity was discussed in Ref.~\cite{Wen:1992uk} in the case of the topological $BF$ theory for Abelian anyons, and we extend it to the case of our generalized setup. 

We may work on another basis of gauge fields $\widetilde{a}_A$, $\widetilde{b}_i$, which are related to the original fields as  
\be
a_A = M_{AB} \, \widetilde{a}_B, 
\quad 
b_i = \widetilde{b}_j \, N_{ji} , 
\label{eq:basis}
\ee
with some invertible matrices $M$ and $N$. 
Under the basis change, the $K$ matrix  should be replaced as
\be
\widetilde{K}= N K M. 
\ee
One should not forget that, together with Eq.~(\ref{eq:basis}), 
we also have to transform the charge matrices, 
\be
\widetilde{Q} = QM, 
\quad
\widetilde{R} = N R . 
\ee
Under this change of normalization, the coefficients of kinetic terms are also changed as 
\be
\widetilde{G}^b=N G^b N^T,\; \widetilde{G}^a=M^T G^a M. 
\ee 
%
%
The theories connected by this transformation are equivalent and 
give the same physical results. 
For example, gauge invariance of the theory is not affected by the basis change, 
since the factor $R^{-1} K Q^{-1}$ appearing in Eq.~(\ref{eq:gauge-delta}) is invariant under the basis change, Eq.~(\ref{eq:basis}):
\be
\widetilde{R}^{-1} \widetilde{K} \widetilde{Q}^{-1}=R^{-1}K Q^{-1}.
\ee

One should be cautious in changing the basis because 
it in general introduce the coupling between massive and massless 
gauge fields in the kinetic terms, which complicates the analysis. 
One may seek a simpler expression of $K$ matrix by changing the basis\footnote{
A similar comment is true also for 3d Chern-Simons theory with a slightly different reasoning~\cite{Wen:1992uk}.}; 
for example, we can always obtain one of the simplest expressions for  $K$ matrix by considering the Smith normal form\footnote{Authors appreciate the anonymous referee for pointing this out. }
\be
K= U^{-1} K' V^{-1}, 
\ee
where $U$ and $V$ are integer matrices with $\det(U)=\det(V)=\pm1$ and $K'$ takes the form 
\be
K'=\begin{pmatrix}
d_1&0&0&\ldots\\
0&d_2&0&\ldots\\
0&0&\ddots&\\
\vdots & \vdots &&
\end{pmatrix}
\ee 
where $d_i$ are also integers. 
Since $\det(U)=\det(V)=\pm 1$, we can stay in the canonical normalization of gauge fields when the original expression is in the canonical normalization. 
If (\ref{eq:assumption_orthogonality}) is maintained, this is the simplest basis to work on. 
However, this is not always the case. 
The theories considered in Secs.~\ref{sec:cfl} and \ref{sec:tos} are such examples.  
 
Generically, 
if the basis change matrices satisfy $N N^T = 1$ and $M^T M =1$, 
the condition (\ref{eq:assumption_orthogonality}) is maintained under the basis change. 
In this case, the Moore-Penrose inverses of old and new $K$ matrices 
are related by\footnote{When $K$ is invertible,  Eq.~(\ref{eq:k+k+}) is trivially true. }
\be
\widetilde K^+ = M^{-1} K^+ N^{-1} . 
\label{eq:k+k+}
\ee
Otherwise, 
the condition (\ref{eq:assumption_orthogonality}) is not kept and 
Eq.~(\ref{eq:k+k+}) does not hold. 
In such cases, 
the change of basis introduces the coupling between massless and massive modes, 
which complicates the evaluation of correlation functions and 
identification of symmetry generators. 
For a detailed discussion on the effects of 
the coupling of massless and massive modes, 
see Appendix~\ref{sec:smith}.

\section{Higher-form symmetries of generalized $BF$ theories}\label{sec:emergent}

Let us examine the global higher-form symmetry of the action
(\ref{eq:seff}). 
Depending on the structure of the $K$ matrix, 
there can exist discrete 1-form and 2-form symmetries. 
In the presence of massless phonons and photons, 
there are also continuous 1-form and 2-form symmetries. 
Here let us identify those symmetries. 

\subsection{Continuous higher-form symmetries}

The action (\ref{eq:seff}) has 
continuous (or $U(1)$) 1-form or 2-form global symmetries 
in the presence of massless photons or NG bosons, respectively. 
The symmetry transformation is given by 
\begin{equation}
 a_A \mapsto a_A +\epsilon\, C^{\alpha}_A \, \mu_\alpha, 
  \quad
 b_i \mapsto b_i +\epsilon\, D^{\bar\alpha}_i \, \rho_{\bar\alpha}, 
\label{eq:continuous_higher_form}
\end{equation}
where $\mu_\alpha$ and $\rho_{\bar\alpha}$ are flat 1-form and 2-form
fields and $\epsilon$ is an infinitesimal parameter. 
One can immediately see that the effective action (\ref{eq:seff}) is invariant under the transformation 
(\ref{eq:continuous_higher_form}), 
using $K_{iA} C^\alpha_A = 0$ and $D_i^{\bar\alpha} K_{iA}=0$. 
The number of $U(1)$ $1$-form symmetry is given by 
$|\alpha|=\dim ({\rm ker}\, K)$, 
and the number of $U(1)$ $2$-form symmetries is 
$|\bar\alpha|= \dim ({\rm coker}\, K)$.
They are the same as the number of massless photons and massless NG modes. 
Those symmetry act on Wilson loops and vortex operators as a phase rotation, 
\be
W_A \mapsto 
W_A \exp \left( 2\pi \im  \epsilon \, C^{\alpha}_A \, \mu_{\alpha} \right), 
\quad 
V_i \mapsto 
V_i \exp \left( 2\pi \im  \epsilon \, D^{\bar\alpha}_i \rho_{\bar\alpha}  \right). 
\ee

\subsection{Discrete higher-form symmetries}\label{sec:discrete_higher_form}

The theory can have discrete higher-form symmetries. 
Consider the following transformation of the 2-form gauge field $b_i$, 
\begin{equation}
 b^T_i\mapsto
  b^T_i + [q^T Q K^+]_{i} \,  \lm , 
  \label{eq:bt}
\end{equation}
where 
$\lm = \lm (x)$ is a closed 2-form connection, i.e. $\diff \lambda=0$, 
with quantized holonomy $\int_S \lm \in 2 \pi \mathbb Z$ 
over a closed surface $S$, 
and 
$q_{A}$ is a charge vector. 
In order to make a one-to-one correspondence between $q_A$ and this transformation, we require that the charge vector has to satisfy   
\be
q^T \cdot Q\in ({\rm coker}\, K^+)^\perp=({\rm ker} \, K)^\perp . 
\ee 
The kinetic term for $b_0$ is invariant under  (\ref{eq:bt}),
since it does not shift 
massless NG modes ($\delta [P^b_{ij}b_i] = 0$ 
follows from the definition of the Moore-Penrose inverse). 
The variation of the action is 
\begin{equation}
 \delta_q S_{\rm eff}
  =\frac{\im}{2\pi} [q^T Q K^+ K]_{A B} \int \lm \wedge \rd a_B . 
\end{equation}
Noting that $K^+ K$ is the projection matrix to $({\rm ker} \, K)^\perp$, 
\be
(q^T Q)_A  [K^+ K]_{AB} = (q^T Q)_B. 
\ee
 Thus, 
\begin{equation}
 \delta_q S_{\rm eff}
  =2\pi \im (q^T Q)_A  \int {\lm\over 2\pi} \wedge {\rd a_A\over 2\pi}  \in \sum_A 2 \pi \im  q_A \mathbb Z , 
\end{equation}
where we used the normalization condition for $a_A$. 
In order for this to generate symmetry, we must require $\delta_q S_{\mathrm{eff}}\in 2\pi \im \mathbb{Z}$, and then $q_A$ has to be an integer-valued vector. 

Another necessary requirement of symmetry is that there must exist a physical operator with nontrivial transformation. 
In this case, if vortex operators in (\ref{eq:wvqr}) transforms nontrivially under (\ref{eq:bt}), 
it is a $2$-form symmetry of the system. 
The action of the discrete $2$-form transformation (\ref{eq:bt}) is given by 
a phase rotation, 
\be
V_i^{(R)} 
\mapsto 
V_i ^{(R)}
\exp 
\left(
2\pi \im  \, [q^T Q K^+ R]_{i} 
\right). 
\label{eq:vi-tr}
\ee
If there exists a charge vector $q$ with $q^T Q\in ({\mathrm{coker} \, K^+})^{\perp}$ such that the 
transformation (\ref{eq:vi-tr}) is nontrivial, then 
the system has a discrete $2$-form symmetry. 

Similarly, the discrete 1-form symmetry transformation is given by 
\begin{equation}
 a_A \mapsto a_A + [K^+ R\,  p]_{A}\, \omega ,
\label{eq:1form_trans}
\end{equation}
where $\omega$ is a flat connection with $\int \omega \in 2 \pi \mathbb Z$, 
and $Rp  \in ({\rm coker}\, K)^\perp$. 
The action is varied as 
\begin{equation}
 \delta_p S_{\rm eff}
  =  \frac{\im}{2\pi}[K K^+ R\, p]_{i} \int  \rd b_i \wedge \omega  ,
\end{equation}
Noting that $K K^+R \, p = R\, p$, 
we have $\delta S_{\rm eff} \in 2\pi  \im \mathbb Z$ when $p$ is an integer vector, 
and the system is invariant under the $1$-form transformation.
On the gauge-invariant Wilson loops in (\ref{eq:wvqr}), it acts as 
\be
W_A^{(Q)} \mapsto 
W_A^{(Q)} 
\exp 
\left(
2\pi \im \, [ Q K^+ R\, p]_A 
\right). 
\label{eq:wa-tr}
\ee
If there is a charge vector $p_i$ such that Eq.~(\ref{eq:wa-tr}) is a nontrivial 
transformation, this is a discrete $1$-form symmetry.

\subsection{Particle-vortex statistics}\label{sec:fractional}

Let us discuss the nature of particle-vortex statistics. 
Test particle of charge $q$ and vortex with charge $p$ are represented by
\begin{equation}
 W_{q}^{(Q)}(C) = \exp \left( \ri \, (q^T Q)_A \int_C a_A\right),
  \quad
 V_{p}^{(R)}(S) = \exp \left( \ri \,  \int_S b_i\, (R\, p)_i\right),  
\end{equation}
respectively, and the gauge invariance requires that $q$ and $p$ are integer-valued vectors. Here, $C$ denotes the world-line of the test particle $q$, and $S$ denotes the world-sheet of the vortex $p$. 
As shown in Appendix \ref{app:derivation}, the correlation function of these operators in the effective theory (\ref{eq:seff}) satisfies\footnote{
When the orthogonality condition is not satisfied and 
there are mixed kinetic terms of massless and massive modes, 
there will be corrections, as 
explained in Appendix~\ref{sec:smith}. 
Compare this equation with Eq.~(\ref{eq:app:wv}). 
}
\begin{equation}
\frac{
 \left< W_{q}^{(Q)}(C) V_{p}^{(R)}(S)\right>
 }{  \left< W_{q}^{(Q)}(C)\right> \left< V_{p}^{(R)}(S)\right>}
 = \exp \left[ - 2 \pi \ri \, (q^T QK^+R\, p) \, {\rm Lk}(S, C) \right],
 \label{eq:wv}
\end{equation}
where ${\rm Lk}(S,C)$ is the linking number
of the loop $C$ and the surface $S$. 
This phase $\rme^{2\pi\im (q^TQ K^+R p)}$ gives the mutual statistics between test particles and vortices. 
Here, we intensionally use the same symbol $q,p$ to represent the charges of $W,V$ and to parametrize the discrete higher-form transformations in Sec.~\ref{sec:discrete_higher_form}. We shall soon see that the discrete higher-form symmetries are generated by these operators when $p,q$ are chosen so that $V_p^{(R)}$, $W_q^{(Q)}$ are topological~\cite{Gaiotto:2014kfa}. 

In order to identify the generators of discrete higher-form symmetries, let us discuss when the vortex operator $V_p$ becomes topological. 
This can be seen by deforming the surface as 
\bea
V_{p}^{(R)}(S + \delta S) 
&=& V_{p}^{(R)}(S) 
\exp
\left(
\im (R\, p)_i 
\left\{\int_{S+\delta S} b_i 
- \int_{S} b_i 
\right\}\right)\nonumber\\
&=& 
V_{p}^{(R)}(S) 
\exp
\left(
\im (R\, p)_i 
\int_V \rd b_i 
\right), 
\eea
where $V$ is the volume swept by the continuous deformation $S \rightarrow S+\delta S$, or $\delta S=\p V$. Therefore, the operator $V^{(R)}_p$ is topological, i.e. $V_{p}^{(R)}(S+\delta S)  =V_{p}^{(R)}(S)$, if 
\be
(R\, p)_i \diff b_i=0 , 
\label{eq:eom_pb}
\ee
by using the EOM of (\ref{eq:seff}). 
This happens when the vortex decouples from the massless phonon, i.e. $R\, p\in ({\rm coker}\, K)^{\perp}$. 
Indeed, the EOM of (\ref{eq:seff}) with respect to $a_A$ gives 
\be
(\diff \star \diff a^T)G^a (1-K^+ K)+{\im\over 2\pi}\diff b^T K=0. 
\ee
Multiplying $K^+ R\,p$ from the right and using $KK^+ R\, p=R\, p$ for $R\, p\in ({\rm coker} K)^{\perp}$, we get (\ref{eq:eom_pb}). 
Other vortex operators with $R\, p\not \in ({\rm coker}\, K)^{\perp}$ show non-topological behaviors because of its coupling to massless NG excitations. 
Let us now discuss the physical meaning of the expectation value (\ref{eq:wv}) when we consider the topological surface operator $V_p^{(R)}$ with $R\, p\in ({\rm coker}\, K)^{\perp}$. 
Taking $S$ as a two sphere singly linked to $C$, we get
\be
 \left< W^{(Q)}_{q}(C) V^{(R)}_{p}(S)\right>
 = \rme^{2\pi \im (q^T QK^+ R\, p)}\left< W_{q}(C)\right>,
\ee
because $\langle V^{(R)}_p(S)\rangle =1$ as $S$ can be shrunk to a point. 
This explicitly shows that the insertion of $V_p^{(R)}$ with $R\, p\in ({\rm coker} K)^{\perp}$ is nothing but the $1$-form transformation~(\ref{eq:1form_trans}). 

The same discussion applies for the Wilson loop. When $q^T Q\not\in ({\rm ker}\, K)^{\perp}$, the Wilson loop obeys Coulomb law.  $W_q^{(Q)}(C)$ shows topological dependence on $C$ if and only if 
\be
(q^T Q)_A \diff a_A=0, 
\ee
by the EOM of (\ref{eq:seff}), which is equivalent to the decoupling condition to massless photons, i.e. $q^T Q\in ({\rm ker}\, K)^{\perp}$. 
When $q^T Q\in ({\rm ker}\, K)^{\perp}$, the expectation value (\ref{eq:wv}) can be written as 
\be
 \left< W_{q}^{(Q)}(C) V_{p}^{(R)}(S)\right>
 = \rme^{2\pi \im (q^T Q K^+R\, p)}\left< V_{p}^{(R)}(S)\right>,
\ee
because $\langle W_q^{(Q)}(C)\rangle =1$ by its topological nature. This is nothing but the $2$-form transformation (\ref{eq:bt}) if the phase $\rme^{2\pi \im (q^T QK^+ Rp)}$ is nontrivial for some vortex charge $p$.

Note that the braiding phase $\e^{\im(q^TQ K^+ R\, p)}$ is invariant under 
\be
(R\,p)_i \mapsto  (R\,p)_i + [K  \Lambda]_i , 
\ee
where $\Lambda_A\in ({\rm ker} \,K)^{\perp}$ is an integer vector. 
Two charge vectors related by this relation are 
topologically equivalent. 
Likewise, $\e^{\im(q^T Q K^+ R\,p)}$ is invariant under 
\be
(q^T Q)_A \mapsto  (q^T Q)_A +  [(\Lambda')^T K ]_A , 
\ee
with an integer vector $\Lambda'_i \in ({\rm coker}\, K)^{\perp}$.

\subsection{Fate of the symmetries and topological order } 

Let us discuss the fate of symmetries. 
For notational simplicity, let us consider the case $Q=1$ and $R=1$ in this subsection. 
The result for general cases can be reproduced by replacing $(q^T K^+ p)$ by $(q^T Q K^+ R p)$. 

When discrete higher-form symmetries are spontaneously broken, 
there appear deconfined anyons and the system acquires a topological order. 
Let us diagnose the existence of topological order in the current theory. 
The charged object of a discrete $2$-form symmetry is 
a vortex operator with a vortex charge $p$. 
Existence of a discrete 2-form symmetry indicates that 
there exists a topological Wilson loop. 
A Wilson loop with charge $q\in ({\rm ker}\, K)^{\perp}$ induces a $2$-form transformation on 
a vortex of the form, 
\be
V_{p} \mapsto V_{p} \exp [2\pi \im \, q^T K^+ p] . 
\ee
A discrete $2$-form symmetry is spontaneously broken if 
the expectation value of the vortex operator obeys the surface law 
at large vortex world-surface $S$, 
\be
\langle
V_{p} (S)
\rangle \simeq \exp (- \kappa \, {\rm perimeter}[S]) ). 
\ee

If a vortex operator is charged under a $U(1)$ 2-form symmetry, 
$\langle V_p(S)\rangle$ decays faster than the 
surface law, meaning that the $2$-form symmetry is unbroken. 
This is because, in $4$ spacetime dimensions, 
$2$-form continuous symmetry cannot be broken 
due to the generalized Coleman-Mermin-Wagner theorem for higher-form
symmetry \cite{Gaiotto:2014kfa,Lake:2018dqm}. 
Physically, this corresponds to the fact that 
vortices couple to massless NG bosons and feel logarithmic confining force. 
In order for the system to have a topological order, 
there has to be an excitation that is neutral
under $U(1)$ $1$-form or $2$-form symmetries. 
A $U(1)$ $2$-form transformation (\ref{eq:continuous_higher_form}) induces a phase rotation of the form 
\be
V_{p} \mapsto V_{p} \exp [2\pi \im  (D^{\bar\alpha}_i \epsilon_{\bar\alpha}) p_i ], 
\ee
where $\epsilon_{\bar\alpha}$ is a continuous parameter. 
In order for a vortex to be neutral under this symmetry, 
we have to take the vortex charge as  
$
p \in ({\rm coker}\, K)^\perp , 
$
which is the same condition with the one that $V_p(S)$ is a topological surface operator. 
If there is a generator of discrete $2$-form symmetry specified by $q$ 
that acts nontrivially on such vortices, 
those vortices are deconfined and 
they generate $1$-form symmetries. 
Then the system can have a pair of broken $1$-form and $2$-form symmetries
specified by the pair $(p, q)$.  
In such a case, the system is topologically ordered. 

To summarize the argument above,  
appearance of topological order in the theory (\ref{eq:seff}) can be detected by the following condition:  
\begin{list}{$\clubsuit$}{}  
\item There exists a pair of integer vectors $(p,q)\in ({\rm coker} \, K)^{\perp}\times ({\rm ker}\,  K)^{\perp}$, such that $\rme^{2\pi \im (q^T K^{+} p)}\not =1$
\footnote{
One might think that the former condition 
$(p,q)\in ({\rm coker} \, K)^{\perp}\times ({\rm ker}\,  K)^{\perp}$ 
is redundant, because the components in ${\rm coker}\, K$ or ${\rm ker }\, K$ are 
projected out when they are contracted with $K^+$ and 
we can just project $p$ or $q$ to $({\rm coker}\, K)^\perp$ and 
$({\rm ker}\, K)^\perp$ to obtain topological excitations. 
However, the charge vector after this projection is not necessarily an integer vector, 
so we need the former condition to ensure that topological quasiparticles/vortices 
indeed exist. 
}.
\end{list}

The appearance of the topological order can be explained also by a 't Hooft anomaly~\cite{tHooft:1979rat,  Wen:2013oza,  Kapustin:2014zva, Wang:2014pma} (see also \cite{Witten:2016cio, Tachikawa:2016cha, Gaiotto:2017yup, Tanizaki:2017bam, Komargodski:2017dmc, Komargodski:2017smk, Shimizu:2017asf, Wang:2017loc,Gaiotto:2017tne, Tanizaki:2017qhf, Tanizaki:2017mtm, Guo:2017xex,  Sulejmanpasic:2018upi, Tanizaki:2018xto, Cordova:2018acb, Anber:2018tcj, Anber:2018jdf, Tanizaki:2018wtg, Yonekura:2019vyz} for recent applications),
which is an obstruction in gauging a global symmetry. 
To see this, we introduce background gauge fields 
for a pair of discrete 1-form and 2-form symmetries. 
Suppose that 
$p\in ({\rm coker}\, K)^{\perp}$ generates $\mathbb Z_{N^{(1)}}$ 1-form symmetry and 
$q\in ({\rm ker}\, K)^{\perp}$ generates $\mathbb Z_{N^{(2)}}$ 2-form symmetry. 
The integers $N^{(1)}$ and $N^{(2)}$ are determined as the minimal integers 
so that the following $\Lambda^{(1)}$ and $\Lambda^{(2)}$ are integer vectors, 
\be
\Lambda^{(1)}_A = N^{(1)} [ K^+ p ]_A , 
\quad 
\Lambda^{(2)}_i = N^{(2)} [ q^T K^+ ]_i . 
\ee
We will couple the system to background gauge fields corresponding to those symmetries. 
To do this, it is convenient to write the $BF$ coupling term of the action as 
\be
S_{\rm BF} =\im \frac{K_{iA}}{2\pi} 
\int_{M_5} \rd b_i \wedge \rd a_A , 
\ee
where $M_5$ is a 5-dimensional manifold with $\p M_5 = M_4$. 
$S_{\rm BF}$ defines the four-dimensional local theory because $S_{\rm BF}\in 2\pi \im \mathbb{Z}$ on closed five-dimensional manifolds. 
We can introduce background gauge fields as 
\be
S_{\rm BF, gauged} 
= 
\im \frac{K_{iA}}{2\pi} 
\int_{M_5} 
\left(\rd b_i  + \Lambda^{(2)}_i \mathcal B \right)
\wedge 
\left(
\rd a_A  + \Lambda^{(1)}_A \mathcal A  
\right) , 
\ee
where $\mathcal A$ 
is $\mathbb Z_{N^{(1)}}$ 2-form gauge field 
and 
$\mathcal B$ is a 
$\mathbb Z_{N^{(2)}}$ 
3-form gauge field,
that can be written as 
\be
N^{(1)} \mathcal A = \rd \mathcal A' , 
\quad 
N^{(2)} \mathcal B = \rd \mathcal B' . 
\label{eq:nada-nbdb}
\ee
Since $S_{\rm BF, gauged}$ is obtained by the minimal coupling procedure, it is manifestly invariant under the $\mathbb{Z}_{N^{(1)}}$ 1-form and $\mathbb{Z}_{N^{(2)}}$ 2-form gauge transformations. 
However, $S_{\rm BF, gauged}$ is no longer well-defined as a four-dimensional theory unless $(q^T K^+ p)\in \mathbb{Z}$. 
Indeed, using $q^T K^+ K=q^T$ and $KK^+ p =p$, we find that 
\bea
S_{\rm BF,gauged}&=&2\pi \im\int_{M_5}\left({\diff b^T\over 2\pi}\wedge K {\diff a\over 2\pi}+{N^{(2)}\mathcal B\over 2\pi}\wedge {q^T \diff a\over 2\pi}+{\diff b^T p\over 2\pi}\wedge {N^{(1)}\mathcal A\over 2\pi}\right)\nonumber\\
&&+{2\pi \im}(q^T K^+ p)\int_{M_5}{N^{(2)}\mathcal B\over 2\pi}\wedge {N^{(1)} \mathcal A\over 2\pi}. 
\eea
The first line on the right-hand-side is well-defined as as a local functional on $M_4=\p M_5$ modulo $2\pi \im$, but the second line is not if $q^T K^+ p\not \in \mathbb{Z}$ and is only well-defined modulo $2\pi \im (q^T K^+ p)$. 
This means that the partition function $Z_{M_4}[\mathcal{A},\mathcal{B}]$ of (\ref{eq:seff}) with the background gauge field $\mathcal{A}$ and $\mathcal{B}$ cannot be gauge invariant as a four-dimensional field theory, 
which indicates a 't~Hooft anomaly. 
To make it anomaly free, we need to regard it as a boundary theory of the five-dimensional symmetry-protected topological states, 
\be
Z_{M_4}[\mathcal{A},\mathcal{B}]\exp\left[-2\pi\im (q^T K^+ p)\int_{M_5}{N^{(2)}\mathcal B\over 2\pi}\wedge {N^{(1)} \mathcal A\over 2\pi}\right]. 
\ee
This is an anomaly matching condition by Callan-Harvey's anomaly-inflow mechanism \cite{Callan:1984sa}. 
The existence of a 't Hooft anomaly indicates that the ground state 
cannot be a trivially gapped state. 
In the current case, the anomaly matching condition is satisfied by
the appearance of topological order.

\section{Color-flavor locked phase of QCD}\label{sec:cfl}

As an application of the general framework discussed so far, 
let us discuss the color superconductivity in dense QCD~\cite{Hirono:2018fjr}. 
We will see that there exists the nontrivial mutual statistics between test quark and minimal winding vortices~\cite{Cherman:2018jir}, and 
it indicates the emergence of a $\mathbb{Z}_3$ 2-form symmetry~\cite{Hirono:2018fjr}. 
We show that any vortex operators show algebraic decay instead of the surface law, and thus this 2-form symmetry is unbroken. 
This observation is important to extend the notion of quark-hadron continuity~\cite{Schafer:1998ef} to the continuity as quantum phases at 
zero temperature~\cite{Hirono:2018fjr}. 

We note that this section is the follow up of the previous paper~\cite{Hirono:2018fjr} by the same authors with more detailed presentations. 

\subsection{Generalized $BF$ theory for CFL phase}

A color superconducting phase is realized by the condensation of the Cooper pairs of quarks. 
Let us consider the $3$-color and $3$-flavor QCD with flavor-degenerate mass of fundamental quarks, then the most attractive channel between quarks near the Fermi sea is anti-symmetric in both color and flavor. 
The order parameter field in the effective gauged Ginzburg-Landau description is thus the diquark condensate, 
$\Phi_{\alpha i }$, 
where $\alpha$ and $i$ are indices of the anti-fundamental representation for color and flavor, 
respectively. 
This complex scalar fields $\Phi$ has charge $2$ under $U(1)$ quark number symmetry. 
In the mean-field approximation with an appropriate gauge, we get $\langle \Phi_{\alpha i}\rangle\propto \delta_{\alpha i}$ at sufficiently large quark chemical potentials, and the diagonal transformation of color and flavor is unbroken, 
\be
{SU(3)_\rmc\times SU(3)_\rmf\times U(1)\over \mathbb{Z}_3\times \mathbb{Z}_3}\to {SU(3)_{\rmc+\rmf}\over \mathbb{Z}_3}\times \mathbb{Z}_2. 
\ee
This phase is therefore called color-flavor locking (CFL)~\cite{ Alford:1998mk, Schafer:1998ef, Alford:2007xm}, and there is a massless NG boson associated with the 
spontaneously broken $U(1)$ symmetry. 

Here, let us formally generalize the flavor and color group to be $SU(N)_\rmf$ and $SU(N)_\rmc$\footnote{
A similar gauged GL model with $U(N)_{\rmc}$ gauge group is 
considered in Ref.~\cite{Hidaka:2019jtv}. In this case, all the gauge fields are massive and there is no massless NG modes. 
}. 
The scalar field $\Phi$ is taken to be in the anti-fundamental
representation both for $SU(N)_\rmc$ color group and $SU(N)_\rmf$ flavor group, although it is no longer related to the Cooper pair of fundamental quarks\footnote{Only when $N=3$, the two-index anti-symmetric representation is the same with the anti-fundamental representation. }  when $N\not =3$. 
The effective Lagrangian of the CFL phase is given by 
a gauged Ginzburg-Landau model, 
\begin{equation}
 S={1\over 2g_{\mathrm{YM}}^2}|G|^2
 +{1\over 2}|(\diff + \im a_{SU(N)})\Phi|^2
 +V_{\mathrm{eff}}(\Phi^\dagger \Phi, |\det(\Phi)|),
\end{equation}
where $a_{SU(N)}$ is the $SU(N)_\rmc$ color gauge field, $G$ is its field
strength, and the effective potential $V_{\mathrm{eff}}$ depends only on
the color-singlet order parameters, $\Phi^\dagger \Phi$ and $\det(\Phi)$. 
We here choose $V_\mathrm{eff}$ so that it has the minima at 
\begin{equation}
\Phi^\dagger \Phi=\Delta^2\bm{1}. 
\label{eq:diquark_vev}
\end{equation}
The mean-field approximation then realizes the symmetry breaking pattern of the CFL phase, 
\be
{SU(N)_\rmc\times SU(N)_\rmf\times U(1)\over \mathbb{Z}_N\times \mathbb{Z}_N}\to {SU(N)_{\rmc+\rmf}\over \mathbb{Z}_N}\times \mathbb{Z}_2. 
\ee
Here, we again assign the charge $2$ to $\Phi$ under $U(1)$ symmetry as in the case of $N=3$ QCD, although this is not mandatory for $N\not =3$ because of the absence of its interpretation as Cooper pairs. 

In order to apply the formulation in Sec.~\ref{sec:eft}, we take a gauge fixing so that the diquark field $\Phi$ satisfying (\ref{eq:diquark_vev}) is taken to be diagonal, 
\be
\Phi = \Delta 
\diag 
\left(
\e^{\phi_1}, \cdots, \e^{\phi_N}
\right), 
\ee
where $\phi_i$ are $2\pi$-periodic compact scalar fields to denote the phase fluctuations. 
With this gauge fixing, 
only Abelian part of the gauge fields remain. 
As an example of the Cartan generators of $SU(N)$, let us take the Gell-Mann--type matrices,
\be
\tau_1=\diag(1,-1,0,\ldots,0),\; \tau_2=\diag(1,1,-2,0,\ldots, 0),\ldots, \tau_{N-1}=\diag(1,\ldots,1,-(N-1)). 
\label{eq:Cartan_GM}
\ee
Since we can easily find that 
\be
\exp\left({2\pi \im\over n+1}\tau_n\right)=\exp\left({2\pi\im\over n+1}\tau_{n+1}\right), 
\ee
for $n=1,\ldots, N-2$, $\mathbb{Z}_{n+1}\subset U(1)_{\tau_n}\times U(1)_{\tau_{n+1}}$ does not generate gauge transformations, so the gauge redundancy is represented by 
\be
{U(1)_{\tau_1}\times U(1)_{\tau_2} \cdots\times U(1)_{\tau_{N-1}}\over \mathbb{Z}_{2}\times \cdots\times \mathbb{Z}_{N-1}}\subset SU(N). 
\label{eq:redundancy_GM}
\ee
Let us denote the $U(1)_{\tau_A}$ gauge field as $a_A$. 
The scalar field $\rme^{\im \phi_i}$ has a charge $(\tau_A)_{ii}$ under $U(1)_A$, and thus the charge matrix $K$ of this theory is given by 
a $N \times (N-1)$ matrix, 
\be
K_{iA} = 
\begin{pmatrix}
 1 & 1 & \ldots &1\\
 -1 & 1  & \ldots &1\\
 0 & -2 & \ldots  &1\\
\vdots  & \vdots & \ddots & \vdots \\
 0 & 0 & \ldots& - (N-1)  \\
\end{pmatrix}. 
\label{eq:cfl_K_noncanonical}
\ee
Taking the Abelian duality, we get the effective theory of the form (\ref{eq:EFT_general_03}), 
\be
S={\Delta^{-2}\over 8\pi^2}\int |\diff b_i|^2+{{\rm tr}(\tau_A\tau_B)\over 2 g_{\rm YM}^2}\int \diff a_A\wedge \star \diff a_B+{\im\over 2\pi}\int \diff b^T \wedge K\diff a. 
\label{eq:seff_heavy}
\ee
Although 2-form gauge fields $b_i$ follow the canonical normalization, the $U(1)$ 1-form gauge fields $a_A$ obey the non-canonical normalization 
\be
Q_{AB}\int \diff a_B\in 2\pi \mathbb{Z}, 
\label{eq:cfl_oneform_normalization}
\ee
with $Q_{AB}=K_{(i=A),B}$ for $A,B=1,\ldots,N-1$ because the gauge group is (\ref{eq:redundancy_GM}).

As a result of the diquark condensation, all the gluons become massive by Higgs mechanism and 
there is no massless 1-form gauge field. 
This is equivalent to ${\dim }\, ({\rm ker}\, K) = 0$, and thus we can drop the kinetic term of the gauge field in (\ref{eq:redundancy_GM}) when discussing the low-energy physics. 
On the other hand, 
\be
{\rm coker}\,K=\mathbb{R}\begin{pmatrix}
1\\
\vdots\\
1
\end{pmatrix}, 
\ee
and correspondingly there is one massless NG mode, $ \diff b_0\equiv \diff(b_1+\cdots+b_N)$, 
which is associated with the spontaneous breaking of $U(1)$ baryon number symmetry. 
Thanks to the permutation invariance of the kinetic term of $b_i$, which comes out of $U(N)$ flavor symmetry, the mixed kinetic term between $b_0$ and massive modes does not arise, and (\ref{eq:assumption_orthogonality}) is satisfied. 
Therefore, we can obtain the effective theory of the form (\ref{eq:seff}), 
\be
S_{\rm eff}[b,a]={G^b_0\over 2}\int |\diff b_0|^2+{\im \over 2\pi}\int b^T \wedge K\diff a, 
\label{eq:seff_cfl_noncanonical}
\ee
with $G^b_0=\Delta^2/4\pi^2 N^2$.

Let us perform the basis change in Sec.~\ref{sec:basis_change} with $M=Q^{-1}$ so that we work on the canonically normalized gauge fields $\widetilde{a}_A$, i.e. $\int \diff \widetilde{a}_A\in 2\pi \mathbb{Z}$:
\be
\widetilde{a}_A=Q_{AB}a_B. 
\ee
The effective action (\ref{eq:seff_cfl_noncanonical}) becomes 
\be
S_{\rm eff}[b,\widetilde{a}]={G^b_0\over 2}\int |\diff b_0|^2+{\im \over 2\pi}\int b^T \wedge \widetilde{K}\diff \widetilde{a}, 
\label{eq:seff_cfl}
\ee
with the new charge matrix $\widetilde{K}$, 
\be
\widetilde{K}=KQ^{-1}=
\begin{pmatrix}
 1 & 0 & \ldots &0\\
 0 & 1  & \ldots &0\\
 \vdots & \vdots & \ddots &\vdots\\
0  & 0 & \ldots& 1 \\
 -1 & -1 &\ldots & -1  \\
\end{pmatrix}. 
\ee 
Note that this transformation does not violate the condition (\ref{eq:assumption_orthogonality}) because it only changes the massive gauge fields $a_A$. 
We can directly obtain the effective action (\ref{eq:seff_cfl}) if we use the another Cartan generator, 
\be
\widetilde{\tau}_i=\diag(0,\ldots, 0, \overbrace{1}^{i\mbox{-th}},0,\ldots,0,-1),
\ee
with $i=1,\ldots,N-1$, instead of (\ref{eq:Cartan_GM}). With this Cartan basis, the Abelian subgroup of $SU(N)$ takes the simper form as 
\be
U(1)_{\widetilde{\tau}_1}\times \cdots \times U(1)_{\widetilde{\tau}_{N-1}}\subset SU(N), 
\ee
compared with (\ref{eq:redundancy_GM}). 

We can apply the extra transformation to obtain the Smith normal form, 
\be
\widetilde{K}=U^{-1}\widetilde{K}', 
\ee
with 
\be
U^{-1}=\begin{pmatrix}
 1 & 0 & \ldots &0 &0\\
 0 & 1  & \ldots &0 &0\\
 \vdots & \vdots & \ddots &\vdots &\vdots\\
0  & 0 & \ldots& 1 &0\\
-1 & -1 &\ldots & -1  &1\\
\end{pmatrix},\quad 
\widetilde{K}'=
\begin{pmatrix}
 1 & 0 & \ldots &0\\
 0 & 1  & \ldots &0\\
 \vdots & \vdots & \ddots &\vdots\\
0  & 0 & \ldots& 1 \\
0 & 0 &\ldots & 0  \\
\end{pmatrix}. 
\ee 
The transformation violates (\ref{eq:assumption_orthogonality}), and the result in Sec.~\ref{sec:emergent} cannot be naively applied due to the mixed kinetic term between the massless and heavy degrees of freedom. 
We therefore do not take the Smith normal form in this section, but put the detailed computation with the Smith normal form in Appendix~\ref{sec:smith}.

\subsection{Emergent higher-form symmetry of CFL phase}

Since the Higgs mechanism makes all the gauge fields massive, the Wilson loops
\be
W_{\widetilde{A}}(C)=\exp\left(\im \int_C\widetilde{a}_A\right), 
\ee
obey the perimeter law. They therefore generate discrete $2$-form transformations on the vortex operators,\footnote{
The vortices with minimal energy in the CFL are called non-Abelian vortices or semi-superfluid vortices \cite{Balachandran:2005ev}. 
See Ref.~\cite{Eto:2013hoa} for a review. 
}
\be
V_i(S)=\exp\left(\im \int_S b_i\right)\mapsto 
\rme^{2\pi \im\widetilde{K}^+_{Ai}} V_i(S). 
\ee
The Moore-Penrose inverse of $\widetilde{K}$ is given by 
\be
\widetilde{K}^+_{Ai}=[QK^+]_{Ai}=\delta_{Ai}-{1\over N}, 
\ee
and thus this is the $\mathbb{Z}_N$ 2-form symmetry~\cite{Hirono:2018fjr}. 
This shows that the test quark and the minimal winding vortex has the $\mathbb{Z}_N$ mutual statistics~\cite{Cherman:2018jir}:
\be
\langle W_{\widetilde{A}}(C) V_i(S)\rangle = \exp\left(-{2\pi\im\over N}{\rm Lk}(C,S)\right) \langle V_i (S)\rangle. 
\ee
Note that this phase rotation is a subgroup of the $U(1)$ 2-form symmetry. 

Let us show that the above $\mathbb{Z}_N$ 2-form symmetry is unbroken~\cite{Hirono:2018fjr}. 
To show it, we have to see that any vortex operators $V_p(S)$ of charge $p$ decay faster than the perimeter (surface) law when  the charge $p$ is nontrivial under $\mathbb{Z}_N$ 2-form symmetry. 
$V_p$ is topological if and only if $p\in ({\rm coker}\, \widetilde{K})^{\perp}$, and those operators are generated by 
\be
V_i(S) V_{i+1}(S)^{-1}. 
\ee
Since these operators are neutral under $\mathbb{Z}_N$, 
we find that this symmetry is unbroken. 
This also means that the theory has no 1-form symmetry since any topological surface operators are neutral with the Wilson loops. 
As a consequence, the CFL phase at the zero temperature acquires the emergent $\mathbb{Z}_N$ 2-form symmetry, but there is no topological order since it is unbroken.

\subsection{Implications for the quark-hadron continuity scenario}

Let us comment on the physical consequences of the unbroken $\mathbb Z_3$ 2-form symmetry 
regarding the quark-hadron continuity scenario \cite{Schafer:1998ef, Hirono:2018fjr}. 
Conventionally, the phases of matter have been classified by 
the (0-form) symmetry of the system. 
If there are two phases with the same symmetry, they are 
considered to be in the same phase. 
It means that there exists a certain deformation of the Hamiltonian 
by which the two phases can be continuously connected.
When we consider quantum phases of matter, 
there can be phases with the same 0-form symmetry but are 
distinguished by different topological orders. 
Here, we have shown here that the CFL phase does not have 
a deconfined discrete gauge field and 
that implies the absence of topological order, 
although the appearance of fractional braiding phase has a certain similarity 
to the nature of a topological ordered state. 
The braiding phase is shown to be a direct consequence of a 
(unbroken) $\mathbb Z_3$ 2-form symmetry. 
In addition, the system does not have a 1-form symmetry, and 
hence there is not mixed 't~Hooft anomaly of discrete higher-form symmetries, 
which allows for a topologically trivial ground state. 
So, as far as the ground state property is concerned, 
the CFL can be continuously connected to a nucleon superfluid phase,
which presumably has a trivial topological structure
because of the absence of deconfined gluons. 
Thus, we have extended the continuity scenario to zero temperature.  

Note that this does not necessarily mean that 
there is no phase transition between nucleon superfluidity and 
the CFL phase as a function of baryon chemical potential $\mu_{\rm B}$. 
Even if two phases have the same symmetry, 
as liquid and vapor phases of water, there can be a phase transition, 
depending on which path one would take in the parameter space. 
The same is true for the quark-hadron continuity at finite temperatures \cite{Schafer:1998ef}. 
In order to predict what would happen along a particular path 
in a parameter space like ($T$, $\mu_{\rm B}$), 
a more detailed approach is necessary. 

One might argue that the existence of the (unbroken) $\mathbb Z_3$ 2-form symmetry 
gives us a distinction of the CFL and nucleon superfluidity. 
However, as we have shown, the discrete $\mathbb Z_3$ symmetry is in fact a 
subgroup of the $U(1)$ 2-form symmetry, 
and this symmetry is associated with the existence of a massless $U(1)$ NG mode. 
Because this mode also exists in a nuclear superfluid phase, 
the continuous 2-form symmetry is also present in this phase. 
Therefore, nucleon superfluidity and the CFL phase have the 
same 0-form and higher-form symmetries.

\section{Example of superfluidity with topological order}\label{sec:tos}

In this section, let us discuss an example of superfluidity with topological order. We consider the generalized $BF$ theory with the action (\ref{eq:seff_cfl_noncanonical}), 
\be
S_{\rm eff}[b,a]={G^b_0\over 2}\int |\diff b_0|^2+{\im \over 2\pi}\int b^T \wedge K\diff a, 
\label{eq:seff_tos}
\ee
with the $K$ matrix (\ref{eq:cfl_K_noncanonical}). 
In the case of the CFL phase, 
the $U(1)$ 1-form gauge fields $a_A$ obey the non-canonical normalization, (\ref{eq:cfl_oneform_normalization}), and no topological order appears. 
In this section, we instead assume the canonical normalization of gauge fields, $\int \diff a_A\in 2\pi \mathbb{Z}$. 

All the Wilson loops $W_A(C)=\exp\left(\im \int_C a_A\right)$ are topological because the theory (\ref{eq:seff_tos}) is in the Higgs phase. 
The gauge-invariant correlation functions are obtained by 
\be
\langle W_{A}(C) V_i(S)\rangle = \exp\left(-2\pi\im K^+_{Ai}{\rm Lk}(C,S)\right) \langle V_i (S)\rangle. 
\ee
The Moore-Penrose inverse of (\ref{eq:cfl_K_noncanonical}) is given by 
\be
K^+=
\begin{pmatrix}
\left(1-\frac{1}2\right) & -\frac{1}2 & 0  & 0 & \ldots &0\\
\left(\frac{1}{2}-\frac{1}{3}\right) & \left(\frac{1}{2}-\frac{1}{3}\right) & -\frac{1}3 & 0 & \ldots &0\\
\left(\frac{1}{3}-\frac{1}{4}\right) & \left(\frac{1}{3}-\frac{1}{4}\right) & \left(\frac{1}{3}-\frac{1}{4}\right) &  -\frac{1}4 &\ldots&0\\
\vdots &  \vdots & \vdots &  \vdots & \ddots & \vdots \\
\frac{1}{N(N-1)} &\frac{1}{N(N-1)}  & \frac{1}{N(N-1)} &\frac{1}{N(N-1)} &\ldots & - \frac{1}{N} \\
\end{pmatrix}, 
\ee
and we can now find the particle-vortex statistics explicitly. 

The set of topological Wilson lines generates $\mathbb{Z}_2\times \cdots \mathbb{Z}_{N-1}\times \mathbb{Z}_N$ 2-form symmetry. 
To see this, let us take the basis of charge vectors $q_n$ of Wilson lines as 
\be
q_1 = \begin{pmatrix}
1 \\
0\\
0\\
\vdots\\
0
\end{pmatrix},
\,\,
q_2 = \begin{pmatrix}
1 \\
1 \\
0\\
\vdots\\
0
\end{pmatrix},\, \ldots, 
\,\,
q_{N-1} = \begin{pmatrix}
1 \\
1 \\
1\\
\vdots\\
1
\end{pmatrix}, 
\ee
that is, $W_{q_n}(C)=\exp\left(\im\int_C(a_1+\cdots+a_{n})\right)$.  
The fractional phase is determined as
\be
q^T_n K^+=\Bigl(\overbrace{-{1\over n+1}, ,\ldots, -{1\over n+1}}^{n+1},\overbrace{0,\ldots, 0}^{N-1-n}\Bigr) \,\bmod 1. 
\ee
Therefore, $W_{q_n}$ generates $\mathbb{Z}_{n+1}$ 2-form symmetry. 

Similarly, we can find the $\mathbb{Z}_2\times \cdots\times \mathbb{Z}_{N-1}$ 1-form symmetry. 
Topological surface operators are given by the vortex charges $p \in ({\rm coker}\, K)^\perp$, and its basis can be chosen as 
\be
p_1 = \begin{pmatrix}
1 \\
-1 \\
0 \\
\vdots\\
0\\
0
\end{pmatrix},
\,\,
p_2 = \begin{pmatrix}
0 \\
1 \\
-1\\
\vdots\\
0\\
0
\end{pmatrix}, \,\ldots,\,\, 
p_{N-1} = \begin{pmatrix}
0 \\
0 \\
0\\
\vdots\\
1\\
-1
\end{pmatrix}, 
\ee
that is, $V_{p_m}(S)=\exp\left(\im \int_S(b_m-b_{m+1})\right)$. 
The topological surface operator $V_{p_m}$ generates $\mathbb{Z}_{m}$ 1-form symmetry for $m=2,\ldots, N-1$. Indeed, this can be found by 
\be
q^T_n K^+ p_m= -{1\over n+1}\delta_{n+1,m}\, \bmod 1, 
\ee
and thus $W_{q_n}$ and $V_{p_{n+1}}$ have the mutual $\mathbb{Z}_{n+1}$ statistics for $n=1,\ldots,N-2$. 
This shows that $\mathbb{Z}_2\times \cdots \times \mathbb{Z}_{N-1}$ 1-form and 2-form symmetries are spontaneously broken, and thus this theory supports a topological order. 

The $\mathbb{Z}_N$ 2-form symmetry generated by $W_{q_{N-1}}$ is unbroken. This fact comes out of the same discussion given in Sec.~\ref{sec:cfl}. 
All the vortex operators charged under $\mathbb{Z}_N$ are coupled to NG boson, and thus those vortices are logarithmically confined. 
In other words, $\mathbb{Z}_N$ 2-form symmetry is a subgroup of $U(1)$ 2-form symmetry defining the vortex winding numbers, and thus the generalized Coleman-Mermin-Wagner theorem prohibits its spontaneous breaking. 

We therefore conclude that the theory describes superfluidity of one NG mode with $\mathbb{Z}_2\times \cdots \times \mathbb{Z}_{N-1}$ topological order.

\section{Summary and outlook}\label{sec:summary}

We have discussed a general effective theory for superfluids with topological order. 
Starting from a gauged GL model, we have derived 
a low-energy gauge theory written in terms of 2-form and 1-form gauge fields. 
The theory has a structure of a $BF$ theory with a non-square $K$ matrix 
that can have a nontrivial kernel/cokernel, 
coupled with massless NG modes corresponding to the breaking of $U(1)$ symmetries. 
Physical spectrum are classified according to the structure of the $K$ matrix. 
We have discussed the symmetry of the theory. 
There can be discrete 1-form and 2-form symmetries as well as 
$U(1)$ 1-form and 2-form symmetries. 
We have shown that the correlation of vortices and quasiparticles 
is written in terms of the topological information of vortex surfaces and quasiparticle  world-lines. 
We have discussed how to identify the presence of topological order, 
which is summarized in the condition $(\clubsuit)$. 
If there is a vortex operator whose average obeys perimeter law, 
the 2-form symmetry is broken, which indicates the topological order. 

As an application of the framework, we have discussed the CFL phase of dense QCD matter. 
We have analyzed the higher-symmetry of the phase and shown that 
color Wilson loops and vortices show fractional braiding 
as a consequence of $\mathbb Z_3$ 2-form symmetry. 
We have shown that the nuclear superfluid phase and the CFL phase have the same symmetry including higher-form ones, 
which extends the quark-hadron continuity scenario to zero temperature. 
We have also discussed an example of superfluidity with topological order 
in Sec.~\ref{sec:tos} and we identified the symmetry and topological order in this system. 

We believe that the framework discussed in this paper would be useful 
in identifying the topological structure 
in systems where topological order and massless modes coexist, 
which include high $T_{\rm c}$ cuprate superconductivity.

\acknowledgments
The work of Y.~T. was supported by JSPS Overseas Fellowship. 
The work of Y.~H. was supported in part by the
Korean Ministry of Education, Science and Technology,
Gyeongsangbuk-do and Pohang City for Independent
Junior Research Groups at the Asia Pacific Center for
Theoretical Physics. 

\appendix

\section{Moore-Penrose inverse\label{app:mp}}

Here we summarize the properties of the Moore-Penrose inverse
that are used in the paper. 
 
Let $K_{iA}$ be a $|i| \times |A|$ matrix with real entries.  
The $|A| \times |i|$ matrix $[K^+]_{Ai}$ that satisfies
the following 4 conditions, 
\begin{eqnarray}
 K K^+ K &=& K ,  \\
 K^+ K K^+ &=& K^+ , \\
 (K K^+)^T &=& K K^+ , \\
 (K^+ K)^T &=& K^T K ,
\end{eqnarray}
is called the Moore-Penrose inverse of $K$,
which is a generalization of the inverse matrix.
For any matrix, its Moore-Penrose inverse always
exists and is unique. 
If linear equations
\begin{equation}
 K x = b, 
\end{equation}
have any solution,
it has to take the form 
\begin{equation}
 x = K^+ \, b + \left[\bm 1 - K^+  K \right]w, 
\end{equation}
where $w$ is an arbitrary vector.
Solutions exist if and only if $K K^+ \, b = b$. 
The kernel and cokernel of $K$ and $K^+$ are related as follows, 
\begin{equation}
 {\rm ker \,} K^+ = {\rm coker\,} K, 
\end{equation}
\begin{equation}
 {\rm coker \,} K^+ = {\rm ker\,} K. 
\end{equation}
Using $K^+$ and $K$, we can define orthogonal projection matrices in the following
way:
\begin{eqnarray}
P &\equiv& 1 - K^+ K, \quad  {\rm projection\,\,\,to\,\,\, } {\rm ker}\,
 K , \\
P^c &\equiv& 1 - KK^+ , \quad  {\rm projection\,\,\,to\,\,\, } {\rm coker}\,
 K .
\end{eqnarray}

Given two matrices $K$ and $L$, 
$(K L)^+ = L^+ K^+$ is true if and only if either of the following conditions is satisfied: 
\begin{eqnarray}
K^T K &=& 1  , \\
L L^T &=& 1 , \\
L &=& K^T, \\
\dim (\ker K) =& \dim & ( {\rm coker }  L) = 0. 
\end{eqnarray}

\section{Delta function forms}\label{app:delta}

For $n$-dimensional submanifold $M_n$ on $D$-dimensional manifold,
a $D-n$ dimensional delta function form
supported on $M_n$ is defined by 
\begin{equation}
\int_{M_n} A = \int  A  \wedge \delta^{\perp}(M_n)  , 
\end{equation}
where $A$ is any $n$-form. 

An exterior derivative of a delta function form is given by 
\begin{equation}
 \rd \, \delta^{\perp} (M_n) = (-)^{D-n+1} \delta^{\perp}(\p M_n) , 
\end{equation}
where $\p M_n$ is the boundary of $M_n$. 
The delta function form is odd under flipping of the orientation of $M_n$, 
\begin{equation}
 \delta^{\perp}(- M_n) = - \delta^{\perp}( M_n)    . 
\end{equation}
Let $M_{n}$ and $M_{m}$ be $n$- and $m$-dimensional submanifold of the
total space. 
The intersection of those two manifolds, 
$M_{n} \cap M_{m}$, is
$D-n -m$-dimensional manifold.
The delta function form of $M_{n} \cap M_{m}$ is given by  
\begin{equation}
 \delta^\perp (M_{n} \cap M_{m} )
  =
 \delta^\perp (M_{n}) \wedge   \delta^\perp (M_{m}) . 
\end{equation}
When $n + m = D$, the interaction regions are points.
The number of intersection is counted by 
\begin{equation}
 {\rm I} (M_{n}, M_{m}) 
 = 
 \int \delta^{\perp}(M_{n})\wedge \delta^\perp (M_{m}), 
\end{equation}
which is an integer.

\section{Derivation of the braiding phase (\ref{eq:wv}) \label{app:derivation} } 

Here let us derive Eq. ~(\ref{eq:wv}). 
We would like to compute the path integral,
\begin{equation}
\frac{1}{Z}
 \int \Dcal a \Dcal b
  \exp
  \left(
   - S_{\rm eff}
   + \ri \int (Rp)_i\, b_i \wedge j_v + \ri \int (q^T Q)_A\, a_A
  \wedge j_p
  \right), 
\end{equation}
where 
$S_{\rm eff}$ is given in Eq.~(\ref{eq:seff}), 
$j_p$ is a quasiparticle current, and
$j_v$ is a vortex current. 
The sources are represented by 
\begin{equation}
 j_v = \delta^{\perp}(S) ,
  \quad
 j_p = \delta^{\perp}(C) ,  
\end{equation}
where 
$C$ and $S$ are world-line and world-sheet of 
the particle and vortex, 
and 
$\delta^{\perp}(S)$ ($\delta^{\perp}(C)$)
are the 2-form (1-form) valued delta function whose
support is $S$ ($C$). 

By integrating out the fields,
the path integral localized around the solutions
of classical equations of motions. 
Note that the kinetic terms of $a_0$ and $b_0$ can expressed as 
\begin{equation}
 \frac{1}{2} \int G^a_{AB} \, \rd (a_0)_A \wedge \star \, \rd (a_0)_A
  =
  \frac{1}{2}
  \int  [P^a G^a P^a]_{AB} \,\, \rd a_A \wedge \star \rd a_B , 
\end{equation}
\begin{equation}
 \frac{1}{2} \int G^b_{ij}\, \rd (b_0)_i \wedge \star \, \rd (b_0)_i
  =
  \frac{1}{2}
  \int  [P^bG^bP^b]_{ij} \, \rd b_i \wedge \star \, \rd b_j . 
\end{equation}
Integrating out $b$, we obtain 
\begin{equation}
 \frac{1} {2\pi} K_{iA} \, \rd a_A
  +(Rp)_i \, j_v 
- [P^bG^bP^b]_{ij} \, \rd \star \rd b_j
= 0 . 
\end{equation}
Multiplying $K^T$ from left, 
\begin{equation}
 \frac{1} {2\pi}
  [K^T K]_{BA}
  \, \rd a_A
  + [K^T R p]_B 
  \, j_v  = 0, 
\end{equation}
where we used a property of Moore-Penrose inverse, 
$K^T P^b = K^T - K^T K K^+ = 0$. 
We can solve this for $a_A$ as 
\begin{equation}
 a_A = -2\pi [K^+ R p]_{A} \,\rd^{-1} j_v
  + (a_0)_A  ,
   \label{eq:sol-a}  
\end{equation}
up to a closed form. 
Here, we used $[K^T K]^+ K^T = K^+ $, 
and
the contribution from massless components $(a_0)_A$ is
to be determined by using the EOM for $(a_0)_A$. 
$[K^+R p]_{A}$ is in general a fractional number,
and this represents a fractional statistics of vortices and
quasiparticles. 
Likewise, integration of $a_A$ leads to 
\begin{equation}
  \frac{1}{2\pi} \rd b_i K_{iA} + (q^T Q)_A \, j_p
  - [P^a G^a P^a]_{AB}\, \rd \star \rd a_B  = 0, 
\end{equation}
which can be solved for $b_i$ as 
\begin{equation}
 b_i =- 2\pi [q^T Q K^+]_{i} \,\rd^{-1} j_p
  + (b_0)_i  . 
   \label{eq:sol-b}
\end{equation}

Using Eqs.~(\ref{eq:sol-b}) and (\ref{eq:sol-a}) , 
we can evaluate the following terms as
\begin{equation}
\begin{split}
 &S_{\rm BF}
 +(R p)_i \int b_i \wedge j_v + (q^T Q)_A \int a_A \wedge j_p\\
 &= - 2 \pi (q^T Q K^+ R p) \int \rd^{-1} j_v \wedge j_p
 + (Rp)_i (b_0)_i \wedge j_v + (q^T Q)_A (a_0)_A \wedge j_p
\\
 &=
-  2 \pi (q^T Q K^+R p) {\rm Lk}(S, C)
 +(R p)_i (b_0)_i  \wedge j_v + (q^T Q)_A (a_0)_A \wedge j_p 
\end{split}
\label{eq:lk}
\end{equation}
Here, we used 
\begin{equation}
 \rd^{-1} \delta^\perp (C) =- \delta^{\perp} (D), 
 \quad
 \rd^{-1} \delta^\perp (S) = \delta^{\perp} (V),
\end{equation}
where $D$ and $V$ are chosen so that $C$ and $S$ are their boundaries 
($C = \p D $ and $S = \p V$),
and 
\begin{equation}
 \int \delta^{\perp}(V) \wedge \delta^{\perp}(C)
  = {\rm I}(V,C)
  = {\rm Lk}(S,C) 
\end{equation}
where ${\rm I}(V,C)$ is the intersection number of $V$ and $C$. 
${\rm I}(V,C)$ is computed by counting the number of intersection with signs
that corresponds to orientation. 

By taking a ratio as in Eq.~(\ref{eq:wv}) 
and using Eq.~(\ref{eq:lk}),
we can see that 
the non-topological contributions related to $(b_0)_i$ and $(a_0)_A$ cancels out,
resulting in Eq.~(\ref{eq:wv}). 
Even though Wilson loops and vortex operators do not show the perimeter
law,
the geometry a Wilson loop does not affect a vortex
and vice versa,
because $(a_0)_A$ do not interact with $(b_0)_i$. 
That is why the ratio (\ref{eq:wv}) is only determined by the topology
of vortices and particle loops.

\section{
Notes regarding basis changes 
}\label{sec:smith}

In this Appendix, let us discuss the influence of basis changes of the gauge fields. 
In particular, when we use a basis 
that does not satisfy the condition (\ref{eq:assumption_orthogonality}), 
the kinetic terms mix the massless and massive modes. 
Although a basis change may simplify the structure of $BF$ terms, 
one should properly take into account the effect of the coupling 
to evaluate correlation functions correctly. 
We first illustrate this point in a simple example, and then discuss more general cases.

\subsection{Example}

We work on the theory given in Sec.~\ref{sec:cfl}. 
 For simplicity, we mainly focus on the case $N=3$. 
The $K$ matrix is given by 
\be
K_{iA} = 
\begin{pmatrix} 
1 & 0  \\
-1 &1  \\
0 & -1  \\
\end{pmatrix} . 
\ee
One can use the Smith normal form to make the $BF$ terms simpler. 
The Smith decomposition of this matrix is given by 
\be
K_{iB}  = U^{-1}_{ij} K'_{jA} V^{-1}_{AB},
\ee
where 
\be
U = 
\begin{pmatrix} 
1 & 0 & 0  \\
1 & 1 &0  \\
1 & 1 & 1  \\
\end{pmatrix}, 
\quad 
K' = 
\begin{pmatrix} 
1 & 0   \\
0 & 1  \\
0 & 0 
\end{pmatrix}, 
\quad
V = \bm 1_2,
\ee
and 
\be
U^{-1} = 
\begin{pmatrix} 
1 & 0 & 0  \\
-1 & 1 &0  \\
0 & -1 & 1  \\
\end{pmatrix}. 
\ee
%
Let us introduce a new basis for 2-form fields by 
\be
c_i = b_j U^{-1}_{ji}. 
\ee
In this basis, $BF$ terms are diagonalized as 
\be
S_{\rm BF} = \frac{\im}{2\pi}
\int
\left[
c_1 \wedge \rd a_1
+ 
c_2  \wedge \rd a_2
\right] . 
\ee
The kinetic term is written as 
\be
S_0 = 
\frac{1}{2} G_{ij} 
\int \rd b_i \wedge \star \, \rd b_j 
= 
\frac{1}{2} 
[U G U^T]_{ij}
\int \rd c_i \wedge \star \, \rd c_j 
\equiv 
\frac{1}{2}  G'_{ij} \int \rd c_i \wedge \star \, \rd c_j  . 
\ee
In the example of the CFL phase, $G_{ij} \propto \delta_{ij}$  
and $G'_{ij} \propto [U U^T]_{ij}$, where 
\be
U
U^T
= 
\begin{pmatrix}
1 &1 & 1 \\
1 & 2 & 2 \\
1 & 2 & 3 \\
\end{pmatrix}. 
\ee
Since $U U^T \neq \bm 1$, 
the orthogonality condition (\ref{eq:assumption_orthogonality}) 
is violated in the new basis. 
%
%
%
The action of $\mathbb Z_3$ symmetry on the rotated basis is 
\be
c_i \mapsto c_i + q_A K^+_{Aj} U^{-1}_{ji} \lambda
\ee
where 
\be
K^+_{Aj} U^{-1}_{ji} 
= 
\begin{pmatrix}
1  & 0 & -\frac{1}3 \\
0  & 1 & -\frac{2}3 \\
\end{pmatrix}
\ee
Thus, this symmetry only acts on $c_3$. 
Below, we compare the computation of $\mathbb{Z}_3$ 2-form symmetry before and after taking the Smith normal form.

\subsection{
$\mathbb Z_3$ symmetry action in the original basis
}

We evaluate the following correlation function, 
\be
\langle
\e^{\im q_2 \int_C a_2}
\, \e^{\im p_3 \int_S b_3}
\rangle . 
\label{app:corr-1}
\ee
First, let us use the original basis. 
We consider 
\be
\begin{split}
S[b_1,b_2,b_3,a_1,a_2]
&= 
 - 
\frac{g^2}{2}
\left(
|\rd b_1|^2 + 
|\rd b_2|^2 + 
|\rd b_3|^2 
\right)
\\
&\quad
- \frac{\im }{2\pi }
\left[
(b_1 - b_2) \rd a_1 
+ 
(b_2 - b_3) \rd a_2
\right]
+ \im  \int p_3 b_3  \wedge \delta^\perp (S)
+ \im \int q_2 a_2  \wedge \delta^\perp (C).  
\end{split}
\ee
The correlation function is calculating by the following quantity 
\be
Z[\delta^\perp (C),  \delta^\perp (S)]
= 
\int 
\prod_i 
 \D b_i 
 \prod_A
 \D a_A
 \exp 
 \left[ 
S[b_1,b_2,b_3,a_1,a_2]
\right] . 
\ee
The integration over $a_1$
gives a delta function enforcing $\rd b_1 = \rd b_2$, 
and also integrating over $b_1$, 
the action is reduced to
\be
S[b_2, b_3, a_2] = 
- \frac{g^2}{2}
\left(
2 |\rd b_2|^2 + 
|\rd b_3|^2 
\right)
- \frac{\im }{2\pi }
\int  (b_2 - b_3) \rd a_2
+ \im \int p_3 b_3  \wedge \delta^\perp (S)
+ \im \int q_2 a_2  \wedge  \delta^\perp (C) . 
\ee
Integration of $a_2$ enforces 
\be
\rd b_2 = \rd b_3 + 2\pi q_2  \, \delta^\perp (C) , 
\ee
and integrating over $b_2$, 
the effective action now becomes 
\be
S[b_3] = 
- 
\frac{g^2}{2}
\left(
2 |\rd b_3 + 2 \pi q_2  \delta^\perp (C)|^2 + 
|\rd b_3|^2 
\right)
+ \im \int p_3 b_3  \wedge \delta^\perp (S) . 
\label{eq:eff-action-b3}
\ee
By performing $b_3$ integration\footnote{
Here and hereafter, we neglect the contributions 
proportional to 
$
\int \delta^\perp (C) \wedge \star  \delta^\perp (C) 
\propto {\rm perimeter}(C), 
 $
 which can be canceled by local counter terms. 
}, 
\be
Z[ \delta^\perp (S),  \delta^\perp (C)] = 
C 
\exp 
\left[
\frac{1}{6 g^2}
(p_3)^2 
( \delta^\perp (S),  (\delta d)^{-1} \star  \delta^\perp (S))
+ 2 \pi \im \frac{2}{3} p_3 q_2 \, {\rm Lk} (C, S)
\right] ,
\ee
where $(x,y) \equiv \int x \wedge \star \, y$ . 
Therefore, we have 
\be
\langle
W_{q_2} (C)
V_{p_3} (S)
\rangle
= 
\e^{2 \pi \im \frac{2}{3} {\rm Lk}(C, S)}
\langle
V_{p_3} (S)
\rangle . 
\ee
This relation means that the Wilson loop $\e^{q_2 \int_C a_2}$ is the generator 
of $\mathbb Z_3$ symmetry. 
The result is unchanged 
if we had projected out the massive kinetic terms in the first place. 
In that case, the kinetic term looks like 
\be
S_0 =- \frac{g^2}{2} \frac{1}{3} |\rd (b_1 + b_2 + b_3)|^2  . 
\ee

\subsection{Calculation in the rotated basis }

Let us evaluate the same correlation function (\ref{app:corr-1})
in a rotated basis that diagonalizes the $BF$ terms. 
Since $c_3 = b_3$, the effective action with sources is written as 
\be
\begin{split}
S[c_1,c_2,c_3,a_1,a_2 ]  
&= 
- 
\frac{g^2}{2}
\left(
|\rd (c_1 + c_2 + c_3)|^2 + 
|\rd (c_2 + c_3)|^2 + 
|\rd c_3|^2 
\right)
\\
& 
\quad
- \frac{\im }{2\pi }
\left[
c_1 \rd a_1 
+ 
c_2 \rd a_2
\right]
+ \im p_3 c_3  \wedge \delta^\perp (S)
+ \im q_2 a_2  \wedge \delta^\perp (C) . 
\end{split}
\ee
Note that, in the rotated basis, there are mixed kinetic terms of massive and massless 2-form fields. 
The field $c_3$ and $a_2$ do not couple in the $BF$ term, 
so naively the Wilson loop of $a_2$ does not seem to generate a phase rotation of 
the vortex operator of $c_3$. 
In order to reproduce the correct results, the coupling of the massive and massless
2-form fields is important. 
Integration over $a_1$ gives $\rd c_1 = 0$, and performing $c_1$ integration, 
\be
S[c_2, c_3, a_2] = 
- 
\frac{g^2}{2}
\left(
2|\rd (c_2 + c_3)|^2 + 
|\rd c_3|^2 
\right)
- \frac{\im }{2\pi }
c_2 \rd a_2
+ \im p_3 c_3  \wedge \delta^\perp (S)
+ \im q_2 a_2  \wedge \delta^\perp (C). 
\ee
Integration over $a_2$ gives 
\be
\frac{\im }{2\pi} \rd c_2 = \im q_2 \delta^\perp (C)  , 
\ee
and after integrating over $c_2$, the resulting action is 
\be
S[c_3 ] = 
- 
\frac{g^2}{2}
\left(
2|\rd c_3 + 2\pi q_2 \delta^\perp (C) |^2 + 
|\rd c_3|^2 
\right)
+ \im \int p_3 c_3  \wedge \delta^\perp (S). 
\ee
This is the same as Eq.~(\ref{eq:eff-action-b3}), 
so we have reproduced the same correlation function. 
Note that the coupling of $c_2$ and $\delta^\perp (C)$ is introduced through the mixed kinetic 
term of $c_2$ and $c_3$. 
If there is no such coupling, 
the Wilson loop of $a_2$ cannot generate $\mathbb Z_3$ phase.

\subsection{More general cases}

Let us consider a more general situation when the matrix $G_{ij}$
specifying the kinetic terms of $b_i$ does not satisfy the 
condition (\ref{eq:assumption_orthogonality}) and 
there are couplings between massless and massive modes. 
We evaluate the correlation function, 
\be
\langle
W_{q} (C) 
V_{p} (S) 
\rangle , 
\label{app:eq:wv-gen}
\ee
with generic charge vectors $q$ and $p$. 
Here we assume that photons $a_A$ are all massive and 
the gauge fields obey the canonical normalization condition, 
namely $Q=\bm 1, R=\bm 1$. 
The effective action is written as 
\be
S[a_A,b_i] =
 - \frac{1}{2}
(\rd b_i , G_{ij} \rd b_j )
 - 
 \frac{\im}{2\pi}
  K_{iA} (b_i, \star^{-1} \rd a_A )
  + \im p_i (b_i , \star^{-1} \delta^\perp (S)) 
  + \im q_A (a_A , \star^{-1} \delta^\perp (C))  .
  \label{app:eq:action-gen}
\ee
Let $P_{ij}$ the projector to ${\rm coker\,} K$. 
We denote a projected index as 
\be
b_{\widehat i} \equiv 
P_{ij} b_j , 
\quad 
b_{\bar i} \equiv (\delta_{ij} - P_{ij}) b_j . 
\ee
Thus, $b_{\hat i}$ denotes massless modes and $b_{\bar i}$ denotes massive modes. 
Since we drop the kinetic terms of massive modes, $G_{\bar i \bar j} = 0$. 
The separability condition (\ref{eq:assumption_orthogonality}) of the massive and massless modes is written as 
\be
G_{i \widehat j} = G_{\widehat i j} . 
\ee
or equivalently 
\be
G_{\bar i \widehat j} =  0  . 
\ee
Below we consider the case when this condition is not necessarily satisfied. 

Now let us compute Eq.~(\ref{app:eq:wv-gen}). 
We evaluate 
\be
Z[ \delta^\perp(C),  \delta^\perp(S)  ]
= \int 
\prod_A \Dcal a_A 
\prod_i \Dcal b_i
\exp S[a_A,b_i] , 
\ee
with the action  (\ref{app:eq:action-gen}). 
The integration over $a_A$ gives 
\be
- \frac{\im }{2\pi }K_{iA} \rd b_i + \im q_A \delta^\perp (C)=0 . 
\label{app:eom-a}
\ee
Let us decompose $b_i$ into massless and massive parts, 
\be
b_i = b_{\wh i } + b_{\bar i } . 
\ee
From Eq.~(\ref{app:eom-a}), 
the massive part is determined (up to closed form) as 
\be
 b_{\bar i }  = - 2 \pi q_A  K^+_{Ai} \, \delta^\perp (S'),
 \label{app:b-massive}
\ee
where $S'$ is a 2-dimensional surface such that 
$\p S' = C$ and 
we used $\delta^\perp (C) = -\rd \delta^\perp (S')$. 
The effective action becomes
\be
S[b_{\wh i}] = 
 - \frac{1}{2}
(b_{\wh i} +b_{\bar i}  , G_{ij}  \delta \rd 
(b_{\wh j} + b_{\bar j}) ) 
  + \im  ( p_{\wh i} b_{\wh i} , \star^{-1} \delta^\perp (S) )
  + \im  ( p_{\bar i} b_{\bar i} , \star^{-1} \delta^\perp (S) ). 
\ee
where $b_{\bar i}$ is given by Eq.~(\ref{app:b-massive}). 
The EOM varying $b_{\wh i}$ is given by 
\be
 G_{\wh ij} \delta \rd  (b_{\wh j} + b_{\bar j})  = \im p_{\wh i} \star  \delta^\perp (S). 
\ee
The massless part is solved as 
\be
b_{\wh i}  = 
\im  G^{-1}_{\wh i \wh j}p_{\wh j} \, ( \delta \rd  )^{-1} \star  \delta^\perp (S)
- G^{-1}_{\wh i \wh j} G_{\wh j \bar k}  b_{\bar k}. 
\ee
The partition function in the presence of particle and vortex sources is now written as 
\bea
\ln Z[\delta^\perp (S),  \delta^\perp (C)] &=& 
- 
\frac{1}{2}
p_{\wh i} \, G^{-1}_{\wh i \wh j}  p_{\wh j}
( \star \delta^\perp (S), (\delta \rd )^{-1} \star  \delta^\perp (S)) \nonumber\\
&&
+ 2\pi \im 
q_A K^+_{A \bar i} 
\left( 
p_{\bar i } - 
G_{\bar i \wh j} 
G^{-1}_{\wh j \wh k} 
p_{\wh k}
\right) 
(\star \delta^\perp (S), \delta^\perp (S')) . 
\eea
Noting that 
\be
(\star \delta^\perp (S), \delta^\perp (S'))  = 
\delta^\perp (S') \wedge \delta^\perp (S) 
= {\rm Lk} (C, S) 
=- {\rm Lk} (S, C) ,
\ee
the correlation function is written as 
\be
\langle
W_{q}(C)
V_{p}(S)
\rangle
= 
\langle
V_{p}(S)
\rangle
\exp
\left[
- 
 2\pi \im 
q_A K^+_{A \bar i} 
\left( 
p_{\bar i } - 
G_{\bar i \wh j} 
G^{-1}_{\wh j \wh k} 
p_{\wh k}
\right) 
{\rm Lk} (S, C) 
\right]. 
\label{eq:app:wv}
\ee
Because of the coupling of massless and massive sector, 
$G_{\bar i \wh j} \neq 0$ and 
there are additional contributions to the term proportional to the linking number. 
Although 
there is no coupling between $a_A$ and $b_{\hat i}$ in the $BF$ terms, 
the Wilson loop can induce a phase rotation of a vortex operator which 
has components $b_{\hat i}$. 

%

\bibliographystyle{utphys}

\bibliography{./QFT,./refs}

\end{document}